\documentclass[11pt]{iopart}
\usepackage[T1]{fontenc}
\usepackage[latin9]{inputenc}

\usepackage[unicode=true,pdfusetitle,
 bookmarks=true,bookmarksnumbered=false,bookmarksopen=false,
 breaklinks=false,pdfborder={0 0 0}, pdfborderstyle={}, backref=false, colorlinks=true]{hyperref}
 \hypersetup{
 citecolor=blue,
 urlcolor=blue,
 linkcolor=blue}
\usepackage[table,svgnames]{xcolor}
\usepackage{array}
\usepackage{units}
\usepackage{amssymb}
\usepackage{graphicx}
\usepackage{float}
\usepackage{stackrel}
\usepackage{babel}

\makeatletter
\AtBeginDocument{\let\LS@rot\@undefined}
\makeatother

\newcommand{\new}[1]{{{#1}}}

\begin{document}

\title{Characterizing spatial point processes by percolation transitions}

\author{Pablo Villegas}
\address{'Enrico Fermi' Research Center (CREF), Via Panisperna 89A, 00184 - Rome, Italy}
\vspace{-0.4cm}
\author{Tommaso Gili}
\address{IMT Institute for Advanced Studies, Piazza San Ponziano 6, 55100 Lucca, Italy.}
\vspace{-0.4cm}
\author{Andrea Gabrielli}
\address{Dipartimento di Ingegneria, Universit\`a Roma Tre, 00146, Rome, Italy}
\address{'Enrico Fermi' Research Center (CREF), Via Panisperna 89A, 00184 - Rome, Italy}
\address{Institute for Complex Systems, Consiglio Nazionale delle Ricerche, UoS Sapienza, 00185 Rome, Italy}
\vspace{-0.4cm}
\author{Guido Caldarelli}
\address{Department of Molecular Sciences and Nanosystems, Ca' Foscari University of Venice, 30172 Venice, Italy}
\address{European Centre for Living Technology, 30124 Venice, Italy}
\address{Institute for Complex Systems, Consiglio Nazionale delle Ricerche, UoS Sapienza, 00185 Rome, Italy}
\address{London Institute for Mathematical Sciences, W1K2XF London, United Kingdom}

\ead{pablo.villegas@cref.it}
\vspace{10pt}
\begin{indented}
\item[] April 2022
\end{indented}

\begin{abstract} A set of discrete individual points located in an embedding continuum space can be seen as percolating or non-percolating, depending on the radius of the discs/spheres associated with each of them. This problem is relevant in theoretical ecology to analyze, e.g., the spatial percolation of a tree species in a tropical forest or a savanna. Here, we revisit the problem of aggregating random points in continuum systems (from $2$ to $6-$dimensional Euclidean spaces) to analyze the nature of the corresponding percolation transition in spatial point processes.  This problem finds a natural description in terms of the canonical ensemble but not in the usual grand-canonical one, customarily employed to describe percolation transitions.  This leads us to analyze the question of ensemble equivalence and study whether the resulting canonical continuum percolation transition shares its universal properties with standard percolation transitions, analyzing diverse homogeneous and heterogeneous spatial point processes.  We, therefore, provide a powerful tool to characterize and classify a vast class of natural point patterns, revealing their fundamental properties based on percolation phase transitions.  \end{abstract}

\vspace{2pc}

\maketitle

Percolation theory, notwithstanding its conceptual simplicity, has proved extremely successful in the description of emergent features in many physical, biological, ecological, and epidemiological problems \cite{broadbent1957,essam1980, Stauffer, Moloney}. Some recent examples of percolation phenomena in diverse fields include epizootics of sylvatic plagues \cite{Davis2008}, viral spread \cite{virus,pandemic}, spatial organization of ecological patterns \cite{Plotkin,Villegas2021}, and long-range coordination in signal transmission among cells \cite{Ojalvo}, to name but a few.  Advances in percolation theory, both theoretical and computational, have been achieved in discrete lattices and networks \cite{Stauffer, Moloney, microcanonical, Newman, networks}. In the simplest case of site percolation, one considers the probability $p$ of occupation for each site in a given lattice/network, and once $p$ overcomes its critical percolation threshold, $p_c$, a giant (percolating) cluster, spanning the whole system emerges.  The phase transition, separating percolating and non-percolating phases, is typically continuous and universal, defining the so-called \emph{standard (or isotropic) percolation} universality class (IP) \cite{broadbent1957,essam1980, Stauffer, Moloney}, which is characterized by a set of well-established critical exponents \cite{Binney, Moloney, wiki}. In some cases, discontinuous or extremely abrupt (``explosive'') percolation transitions have been found \cite{Achlioptas, Ziff-explosive}.  Similarly, dynamical percolation models, including the stochastic dynamics of ``active'' nodes in networks, have proven crucial to understand the propagation of forest fires, epidemics spreading and the dynamics of opinions on networks \cite{Grassberger1983, Cardy1985, Caldarelli2001, Odor2004, Janssen,networks0,networks1,networks2,Radicchi2, LogPot,avalanches,many}.

Despite the success of discrete percolation theory in describing a wide variety of natural phenomena, some others (as various of the examples mentioned above) are more appropriately described in an embedding \emph{continuous} space.  Diverse studies considered ``continuum percolation'' models, in which a variable number of non-interacting and overlapping arbitrarily-shaped objects are randomly placed on, e.g., a two-dimensional space; the percolation threshold has been shown to depend on the shape of the objects \cite{Mertens2012,baker2002,quintanilla2000} as well as on their possible heterogeneity \cite{Consiglio}. In the particular case of discs in a two-dimensional space, the phase transition has been shown to belong to the standard isotropic percolation universality class \cite{gawlinski1981,vicsek1981,hall1985}, a conclusion that is \new{expected} from the perspective of the renormalization group and the universality principle of phase transitions, where microscopic details such as the existence of a discrete lattice spacing are expected to be irrelevant \cite{Binney}.

Let us underline that the analysis of the percolation phase transition in such continuum problems usually relies on a variable number of objects, i.e., it is formulated in the \emph{grand-canonical ensemble}, and the phase transition emerges upon increasing the number of single units or agents.  Studies of fixed-occupancy samples in discrete lattices, i.e., in the {\emph{canonical ensemble}, have been performed in, e.g., \cite{microcanonical, Newman} to compute percolation thresholds and exponents with high precision.  However, thorough studies fixing the total number of units in the system, but considering a continuum spatial embedding are, to the best of our knowledge, still missing. Such problems arise naturally in the context of theoretical ecology as we discuss in what follows. 

For example, continuum percolation setups are of utmost relevance to, e.g., characterize desertification processes, loss of biodiversity, as well as in analyses of complex spatial (vegetation) patterns \cite{Villegas2021, Scanlon, Kefi, Paula, savanna, MoloneyPP, VillegasRS}. In particular, there have been modeling approaches to analyze the spatial distribution of specific tree species in Barro Colorado and Sri Lankan rainforests for which excellent datasets exist \cite{Wiegand2009, Wiegand2007, Velazquez2016}.  For such percolation-type of problems, e.g., to know if a given tree species percolates or not in a tropical forest,
it is not a priori known what the shape of the objects should be. For instance, in the case of trees in a tropical forest, modeled as percolation problem of disks on a continuous two-dimensional space, should one consider the trunk's diameter, the area covered by roots, or the typical radius of seed dispersal?

Thus, studying the abstract percolation problem with a fixed number of units seems natural, assuming they all \new{are circular disc/spheres that} have a typical radius, $r$. Moreover, it is a natural question to ask, what are the critical properties at the percolation transition occurring at some value $r_c$, encountered as $r$ is increased. First, let us emphasize that ensemble equivalence is a well-known property of equilibrium systems \cite{Touchette2015}, but there exist sound non-equilibrium cases where the ensemble equivalence is broken. For example, ensemble non-equivalence has been recently reported in complex networks \cite{squartini2015, squartini2017} and can be of crucial importance to understand non-ergodic systems or systems with emergent ergodicity breaking \cite{Vroylandt2019, Villegas2016}. In particular, different quantities as, e.g. fluctuations and the excess cluster number, exhibit ensemble non-equivalence in continuum percolation problems \cite{Hu2012}. Nonetheless, a delicate and cumbersome relation maps the percolation problem onto a $q\rightarrow1$ Potts model \cite{wu1978}, so it is foreseen that continuum percolation in the canonical ensemble is in the same universality class of standard isotropic percolation. From an even more general perspective, a unified framework addressing aggregation of point patterns and linking the nature of the percolation phase transition they exhibit to the generative process for the points still has yet to be constructed.

The paper is organized as follows. In Section \ref{Theory} the basic concepts of percolation phase transitions needed to analyze the clustering of spatial point processes are briefly presented. In Section \ref{PoissonH}, we analyze the aggregation process of randomly generated spatial point patterns in the continuum, from dimension $d=2$ to $6$. We provide an estimation of the needed critical radius $r$ to generate a percolating cluster, \new{derive} its relation with the filling factor usually given in grand-canonical approaches, and present a unified and straightforward perspective of the scaling relationships illustrating the \emph{ensemble equivalence} and that, as expected, the emergent phase transition in the \emph{canonical ensemble in continuum percolation}, belongs to the isotropic percolation universality class. We therefore analyze in Section \ref{voids} the special case of empty areas. After that, we consider the different emergent universality classes characterizing the most relevant cases of inhomogeneous spatial processes in Section \ref{inhomogeneous1} and the special case of clustered point patterns in Section \ref{clustered}.

\newpage
\section{Aggregation of spatial point processes}\label{Theory}
Our goal is  to characterize the aggregation properties of a fixed number of points, $N$, distributed in a continuum space (from $2$ to $6$ dimensional Euclidean spaces).  For this purpose, we revisit the clustering process proposed in usual continuum percolation problems \cite{gawlinski1981, Plotkin}, which relies on some predefined distance $r$. In particular, one identifies two individual  points as belonging to the same cluster if their Euclidean distance is less than or equal to $r$ \cite{DBSCAN, DBSCAN2}. Thus, a percolating cluster of points exists if a path connects all the points satisfying the previous condition. Figure \ref{Sketch} shows the schematic procedure (each color stands for a different cluster) for small values of the distance parameter $r$, medium distances (where a complex aggregate emerge), and large distances (all the system belongs to a unique cluster). This simple definition allows us to interpret cluster analysis in the language of statistical mechanics and percolation phase transitions \cite{Plotkin}.

\begin{figure}[H]
\centering
\includegraphics[width=0.7\columnwidth]{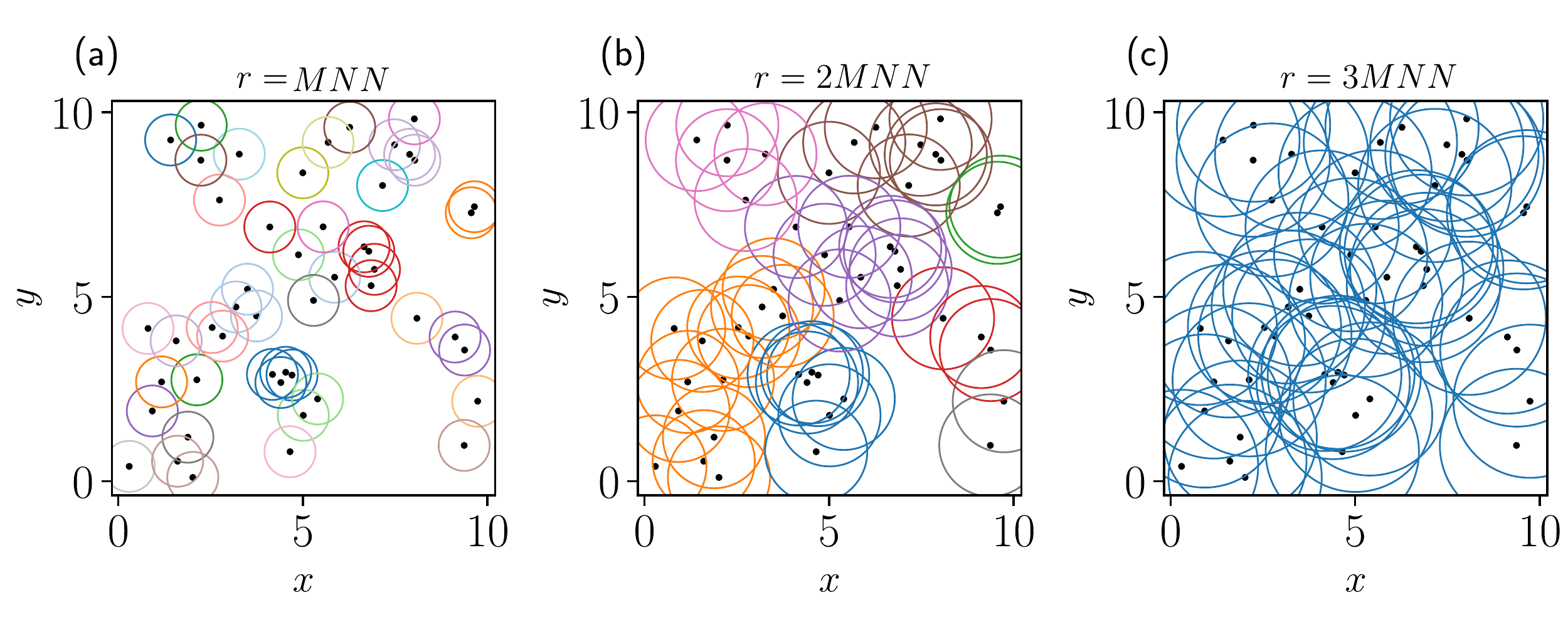}
\caption{Sketch representing the clustering process for different circles of radius $r$, relative to the mean nearest-neighbor distance, $\mbox{MNN}$, of a Poisson point process. The system features a different aggregation of points from (a) isolated clusters to (c) a giant and unique cluster via a critical point (b). Each color represents a different cluster.\label{Sketch}}
\end{figure}

Nearest-neighbor statistics characterize the small-scale structure of such point patterns, determining the typical properties of the distance between any point and its $k^{th}$ nearest neighbor \cite{MoloneyPP}. In particular, the mean \emph{nearest-neighbor} distance ($\mbox{MNN}$), defined as the nearest neighbor mean center-to-center distance between points, makes it possible to assume our processes independent of the considered area while allowing increasing system sizes \cite{Plotkin}. Thus, we normalize the control parameter by the mean nearest-neighbor distance of the point process \cite{GabrielliBook}, i.e., defining $\hat r=r/\mbox{MNN}$, and producing a non-dimensional version of the distance parameter.

Here, we examine the two main variables of percolative systems, i.e., the \emph{cluster strength}, $P_\infty$ and the \emph{mean cluster size}, $\chi$. The infinite cluster (i.e., the maximum cluster) divided by the total number of points, i.e., $P_{\infty}/N$, acts as the order parameter of the system. The distance $\hat r$ acts as the control parameter, showing a percolation phase transition at some critical value $\hat r_{c}$. Alternatively, in usual percolation problems in the grand canonical ensemble, the critical filling factor, $\eta_c$, indicates the percolation transition in the system. This dimensionless quantity is related to the total fraction $\phi$ of the space covered by the objects by $\phi=1-e^{-\eta}$ \cite{Mertens2012}. As a matter of fact --as illustrated in more
detail in \ref{App0}-- it is easy to derive a common relation for hyperspheres in any dimension d, between the critical distance $\hat r_c$, and the critical filling factor, $\eta_c$, as

\begin{equation}
\hat{r}_{c}=d^{2}\frac{\Gamma\left(\frac{d}{2}\right)}{\Gamma\left(\frac{1}{d}\right)\Gamma\left(1+\frac{d}{2}\right)}\eta_{c}^{\nicefrac{1}{d}}
\label{Filling}
\end{equation}
\new{which reduces to $\hat r_c=4\sqrt{\nicefrac{\eta_c}{\pi}´}$ for $d=2$.} 

Turning back to the general discussion, the susceptibility of the system reads,
\begin{equation}
\chi=\frac{\stackrel[i=1]{S}{\sum}s^{2}n\left(s,p\right)}{\stackrel[i=1]{S}{\sum}sn\left(s,p\right)}
\end{equation}
where, as usual, the sum runs over the distribution of clusters of a given size $s$, $n(s,p)$, discarding $P_{\infty}$ if it exists.

At criticality, a set of critical exponents describes the physical behavior near a (continuous) second-order phase transition \cite{Binney}. In particular, we are interested in the relevant quantities,
\begin{center}
\begin{eqnarray}
 P_\infty \propto (\hat{r}-\hat{r}_c)^\beta \\
 \chi \propto (\hat{r}-\hat{r}_c)^\gamma
\end{eqnarray}
\par\end{center}
where $\beta$ and $\gamma$ are the associated critical exponents. Additionally, $D$ represents the fractal dimension of the incipient infinite cluster, and $\nu$ characterizes the divergence of the correlation length, $\xi$.

Similarly, close to the critical point,  the  cluster  size  distribution  assumes  the scaling form,
\begin{equation}
 P(S) = S^{-\tau}\mathcal{F}(S/S_c)
\end{equation}
where $S_c$ is the cutoff due to the finite size of the system, and $\tau$ corresponds to the so-called Fisher-exponent. 

In particular, the finite-size scaling (FSS) ansatz ensures that all the different observables of the system depend on the system size and, in particular, taking into consideration the relation $N=L^d$, the order parameter and the susceptibility follow the FSS relation,
\begin{center}
\begin{eqnarray}
 P_{\infty}(\hat{d}_{c};N) & = & N^{\nicefrac{-\beta}{\nu d}}\mathcal{F}(N^{\nicefrac{1}{\nu}}\hat{d})\\ 
 \chi_{\infty}(\hat{d}_{c};N) & = & N^{\nicefrac{\gamma}{\nu d}}\mathcal{F}(N^{\nicefrac{1}{\nu}}\hat{d}) \label{FSS}
\end{eqnarray}
\par\end{center}
where $\mathcal{F}$ is a scaling function, and $d$ is the physical dimension of the system.
\newpage

Let us remark that the \emph{hyperscaling relation}, in terms of the system size $N$, must be fulfilled,

\begin{equation}
 \frac{\gamma}{\nu}+\frac{2\beta}{\nu}=d
 \label{HS}
\end{equation}

and it holds for any critical system below the upper critical dimension \cite{Binney,Stauffer}. For example, in network' percolation, a similar relationship --that does not account for the physical dimension of the space-- has been reported in terms of the rescaled exponent $\nu'=\nu d$ \cite{Radicchi2010}.

\section{Results}
\subsection{Spatial Poisson point processes} \label{PoissonH}
\subsubsection{Two dimensional case}

The Poisson point process, consisting of $N$ points randomly located in space, constitutes the simplest null model exhibiting statistical independence between individuals. This process is characterized by a homogeneous density $\rho=N/A$, where $A$ is the total area \new{of the system}. Due to computational limitations, we perform the simulations without employing periodic boundary conditions. \new{We highlight that} our particular election does not change our main results, identical \new{in both cases} for large system sizes (see further details in \ref{AppA}).

\begin{center}
\begin{figure}[H]
\begin{centering}
\includegraphics[width=0.8\columnwidth]{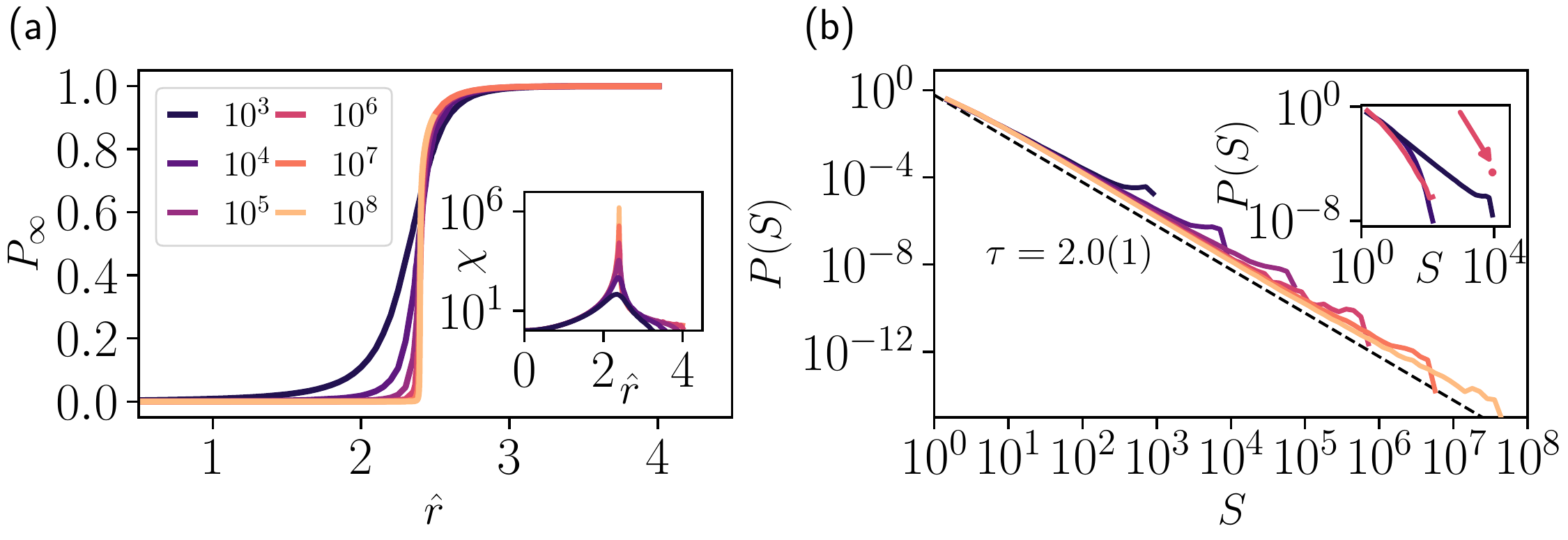}
\par\end{centering}
\caption{\textbf{(a)} $P_{\infty}$ as a function of the distance parameter, $\hat r$, in terms
of the mean nearest-neighbor distance for different number of points
(i.e. system size, see legend). It is possible to observe a clear phase transition at
a critical value $\hat r_c\sim2.39(1)$ scaling with $N$. Inset: Susceptibility
for different system sizes. \textbf{(b)} Cluster size distribution, $P(S)$, versus size, $S$, for different system sizes at the critical value, $\hat r_{c}$. Dashed line shows the exponent $\tau=2.05$. Inset: $P(S)$ versus $S$ for a system size of $N=10^4$ points for a subcritical case ($\hat r = 1.5$), critical case ($\hat r_c = 2.385$) and supercritical case ($\hat r =3.0$). Observe that, for the last case, the probability distribution shows a well-defined bump characteristic of a spanning cluster invading all the system.  Curves have been averaged over $10^{3}-10^{4}$ point patterns. \label{Poiss}}
 
\end{figure}
\par\end{center}

Figure \ref{Poiss}a shows the probability that a given site belongs to the infinite \new{(or largest)} cluster, $P_{\infty}$, versus the normalized distance $\hat r$ (making thus $P_{\infty}$ independent of the area). It exhibits a phase transition at a critical value $\hat r_c \sim2.39(1)$, as we expected from previous works \cite{Plotkin}. Analogous experiments with a variable number of points (see \ref{AppB}), but fixing the radius around them, allowed us to prove that both ensembles, the \emph{canonical} and the \emph{grand-canonical} produce identical results.

Figure \ref{Poiss}b shows the power-law scaling for a two dimensional system, with a characteristic exponent $\tau\sim2.0\left(1\right)$. It is important to mention that converging to the expected $\tau$ value for isotropic percolation ($\tau=2.05$) requires huge sizes (around $10^8$ points, see Fig. \ref{Poiss}b), and thus may lead to interpretation errors. In particular, \new{we observe for $d=2$ that,} the smaller the size the smaller the fitted exponent (see Fig. \ref{Poiss}b, where it converges from $\tau\sim1.75$ for $N=10^3$ to $\tau\sim1.95$ for $N=10^8$). In particular, such convergence does not depend on whether periodic boundary conditions are or not used and it is purely a finite-size effect (see \ref{AppA}). For example, similar behavior has been found in the BTW model for $P(S)$, where the obtained values of the exponents are affected by the finite size of the system \cite{Lubeck1997}.


\begin{center}
\begin{figure}[hbtp]
\begin{centering}
\includegraphics[width=0.6\columnwidth]{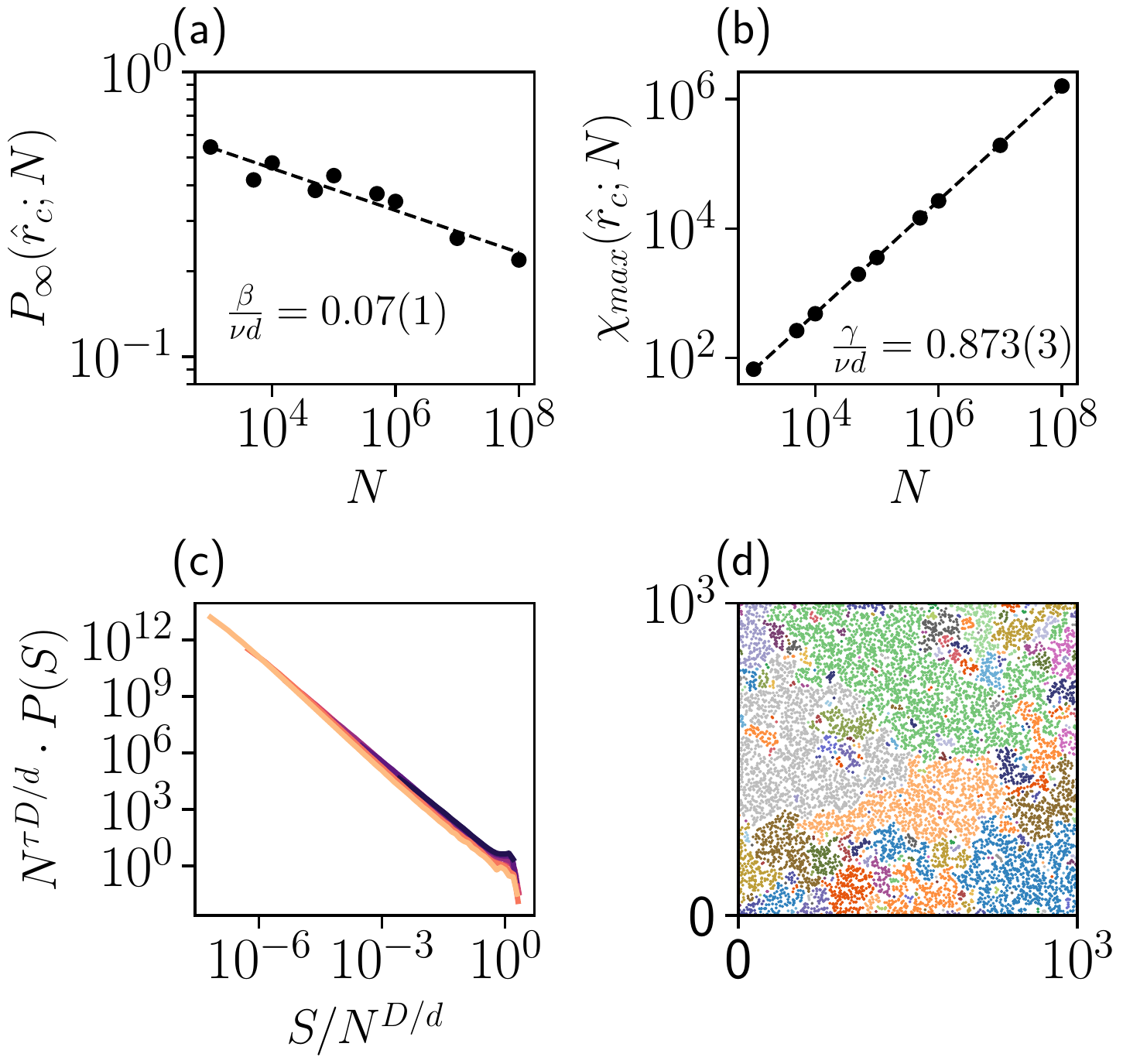}
\par\end{centering}
\caption{\textbf{(a)} Scaling of the order parameter at $\hat r=\hat r_c$, $P_\infty (\hat{r}_c;N) $, versus the system size, $N$. Dashed line represents the best fit to our data to the power law decay $P_\infty(\hat{r}_c;N)\propto N^{-\beta/\nu d}$. \textbf{(b)} Scaling of the susceptibility at $\hat r=\hat r_c$, $\chi (\hat{r}_c;N) $, versus the system size, $N$. Dashed line represents the best fit to our data to the power law behavior $\chi(\hat{r}_c;N)\propto N^{\gamma/\nu d}$. \textbf{(c)} Data collapse analysis of the avalanche size distribution for different system sizes ($N=10^3$ to $N=10^8$). \textbf{(D)} Specific aggregation of a Poisson point pattern of size $N=10^4$ at criticality. Each color stands for a different cluster.  \label{PoissScal}}
 
\end{figure}
\par\end{center}

We computed the finite-size scaling of the different quantities, depending on the system size $N$. In particular, the cluster strength fulfills the relation, $P_\infty (\hat r_c; N) \propto N^{-\beta/\nu d}$, while the average cluster size scales as $\chi(\hat r_c; N) \propto N^{\gamma/\nu d}$ \cite{Moloney,Marro2005}. We can estimate the fractal dimension of the incipient giant cluster through the relation $S_\infty (\hat r_c; N) \propto N^{D/d}$ \cite{Stauffer}. To provide another estimation of the exponent $\tau$ and $D$, we also perform further analyses assuming that the curve $P(S,N)$ can be collapsed into a single one when they are properly rescaled \cite{Chessa1998}. That is, $P_N=\mathcal{F}(S)$ is an universal function under the transformations $P_N=N^{\tau D/d} P(S)$, and $S'=N^{-D/d}S$ \cite{Chessa1998}.

Figure \ref{PoissScal}a shows the FSS of the cluster strength, giving rise to a value $\nicefrac{\beta}{\nu d}=0.07(1)$, while Figure \ref{PoissScal}b shows the system size dependence of the maximum value of the susceptibility, giving the fitted exponent $\nicefrac{\gamma}{\nu d}=0.873(3)$. In addition, Figure \ref{PoissScal}c exhibits the scaling collapse for the cluster size distribution for many system sizes (from $N=10^3$ up to $N=10^8$), which is in agreement with the analysis of $S_\infty(\hat r_c;N)$, and confirms the fitted fractal dimensions $D=1.84(2)$. Fig. \ref{PoissScal}d shows a typical snapshot of different clusters at criticality. The particular set of values lead us to an exponent correction that deserves particular attention; multiplying by the spatial dimension, we obtain the exponents, $\nicefrac{\beta}{\nu}=0.14(2)$ and  $\nicefrac{\gamma}{\nu}=1.75(1)$, belonging to the isotropic percolation universality class.

\subsubsection{Three dimensions and beyond}

The definition of distances between centers (the euclidean distance between them) allows us to generalize our analysis to further dimensions. We computed the same analysis we have done for $d=2$ for dimensions $d=3,4,6$ and $7$. The whole set of critical exponents have been computed using similar finite-size techniques than for the two-dimensional case (see fits, simulations, and scaling collapses for all dimensions in \ref{AppC}).

Figures \ref{PoissDim}(a)-(c) show the probability that a site belongs to the infinite cluster of occupied sites, $P_{\infty}$, versus the normalized distance $\hat r$ for different spatial dimensions. Observe that the higher the spatial dimension, the smaller the value of $r_c$ (in terms of the mean nearest-neighbor distance). Figures \ref{PoissDim}(d)-(e) show the power-law scaling, together with the characteristic exponent $\tau$ for each spatial dimension. Note that, \new{for $d\geq3$,} the obtained exponent present an excellent convergence\new{, for all system sizes,} to the expected $\tau$ value for isotropic percolation depending on the spatial dimension, $d$.

\begin{figure}[hbtp]
\centering
\includegraphics[width=0.8\columnwidth]{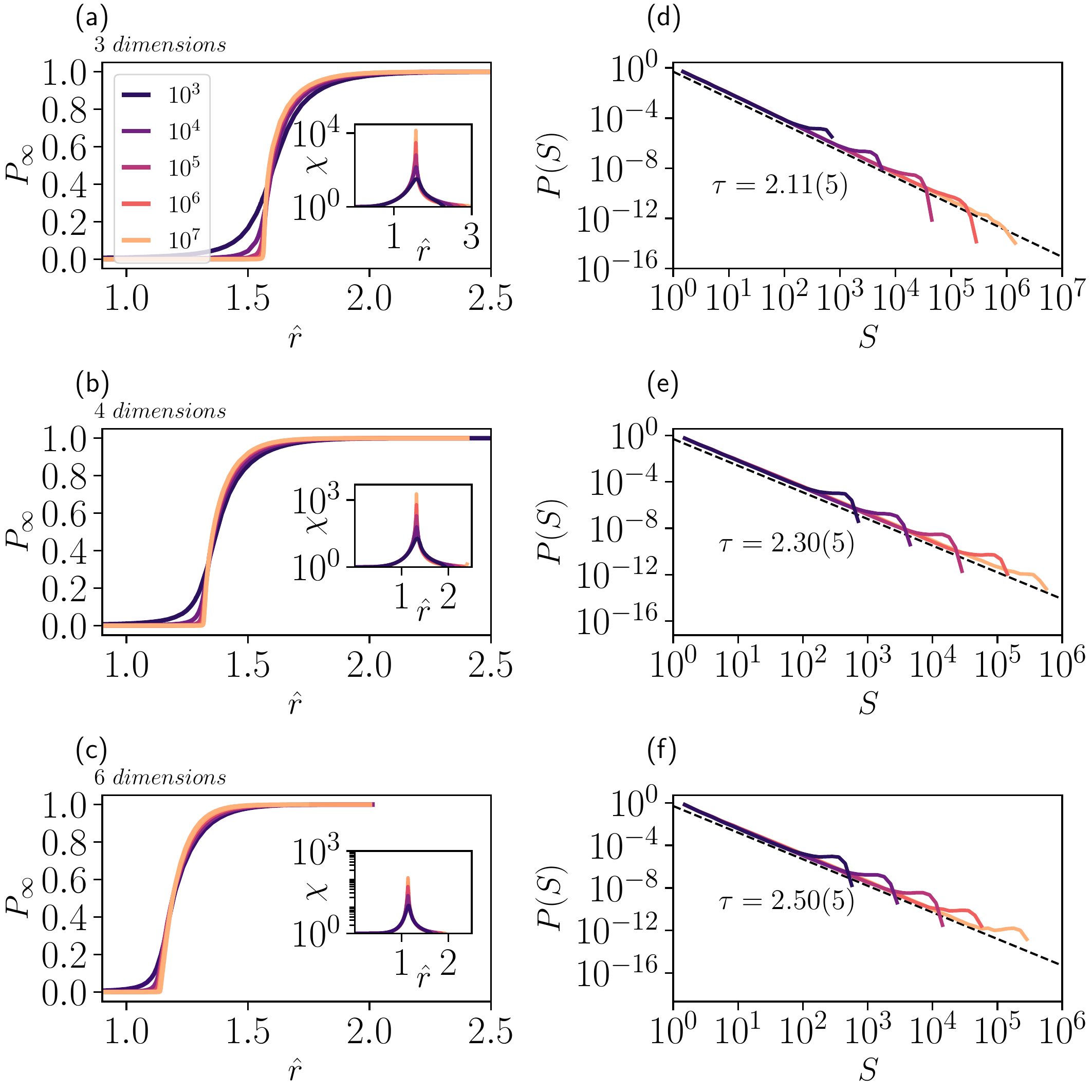}
\caption{\textbf{(a)-(c)} $P_{\infty}$ as a function of the distance parameter, $\hat r$, in terms
of the mean nearest-neighbor distance for different number of points
(i.e. system size, see legend) and different spatial dimensions (see title). It is possible to observe a clear phase transition at
a critical value $\hat r_c$ scaling with $N$, which depends on the spatial dimension of the system. Inset: Susceptibility
for different system sizes. \textbf{(d)-(f)} Cluster size distribution, $P(S)$, versus size, $S$, for different system sizes at the critical value, $\hat r_{c}$. Dashed line shows the best fitted exponent $\tau$ for each spatial dimension.  Curves have been averaged over $10^{3}-10^{4}$
point patterns. \label{PoissDim}} 
\end{figure}

The summary of critical exponents relations and exponents analyzed along the manuscript is shown in Table \ref{dim}, where we have added, for the sake of comparison, some theoretical values in mean-field and two dimensions, and the critical distance from  well-known $\eta_c$ values \cite{Torquato2012} using Eq. \ref{Filling}. Furthermore, as expected at criticality, the hyperscaling relation of Eq. \ref{HS} holds for any dimension lower than the critical dimension. In the same way, the scaling relation $\frac{\gamma}{\nu}+\frac{\beta}{\nu}=D$ holds for all dimensions up to the critical one. From numerical simulations, we observe a critical dimension $d_c=6$, from which the set of critical exponents remains invariant. Observe also that, if the physical dimension of the system is not considered into the scaling forms of Eq.\ref{FSS} (i.e., simply dividing the exponents of Table \ref{dim} by the dimension, $d$), Eq. \ref{HS} is fulfilled in terms of the $\nu'$ exponent, thus giving place to apparent exponents --including the fractal dimension-- with abnormally small values \cite{Radicchi2010}.

\begin{table}[hbtp]
\begin{center}
\begin{tabular}{cccccccc}
\hline 
 Dimension& $\hat r_c$ & $\nicefrac{\beta}{\nu}$ & $\nicefrac{\gamma}{\nu}$ & $\nicefrac{1}{\nu}$ & $D$ & $\tau$ & $\hat r_c (\eta_c)$\tabularnewline
\hline 
\hline 
2 & 2.39(1) & 0.14(2) & 1.75(1) & 0.76(5) & 1.84(2) & 2.0(1) & 2.396906(11) \tabularnewline
\hline 
3 & 1.56(1) & 0.48(3) & 1.97(1) & 1.2(1) & 2.52(3) & 2.11(5) & 1.566161(9) \tabularnewline
\hline 
4 & 1.31(1) & 1.0(1) & 1.98(1) & 1.5(1) & 3.0(1) & 2.25(5)& 1.3260(8) \tabularnewline
\hline 
6+ & 1.13(1) & 1.86(6) & 1.97(3) & 2.2(2) & 4.1(1) & 2.50(5) & 1.15288(21) \tabularnewline
\hline 
2D IP & -- & 0.104 & 1.79 & 0.75  & 1.896 & 2.05 & --\tabularnewline
\hline 
MF IP & -- & 2 & 2 & 2  & 4 & 2.5 & --\tabularnewline
\hline
\end{tabular}

\caption{Summary of the obtained critical exponents relations ($\nicefrac{\beta}{\nu}$ and $\nicefrac{\gamma}{\nu}$) and exponents ($\nu$, $D$ and $\tau$) for different spatial dimensions. The value $\hat r_c$ represents the critical radius (in terms of the mean nearest-neighbor distance) where the percolation transition occurs. We also use the well-known values of filling factors \cite{wikiTh,Torquato2012}, following Eq.(\Ref{Filling}), to validate our estimated values of $\hat r_c$. Finally, the set of mean-field and 2D theoretical relationships and exponents is shown at the end of the table for comparison.}\label{dim}
\end{center}

\end{table}


\subsection{Considerations on the effects of voids} \label{voids}

One of the main concerns to study the macroscopic properties of spatial point process requires to deal adequately with borders \cite{Stoyan1994, hanisch1984}. This issue usually involves two particular problems: knowing the boundaries, how to reduce the biases, and, more subtle, how to correctly identify the not convex borders of a set of points \cite{VillegasRS, AShapesAC}. For instance, establishing proper boundaries is essential to avoid spurious behaviors from the pair correlation function and different spatial density-based measures \cite{VillegasRS}.

Here, we take into consideration the case of a randomly distributed set of points with several excluded areas, as exemplified in the inset of Fig.\ref{Voids}b, for comparison with the simple case of a homogeneous Poisson point process filling all the two-dimensional space. In particular, to analyze the case of the empty areas, we extract for every realization twenty random circles with random radius, $R\in(0.05,0.1)$, distributing all the random points out of them. Figure \ref{Voids}a shows the comparison for the percolation phase transition for a set of points distributed with and without voids, which takes place for identical $\hat r_c$ values. Likewise, just at the critical point, the Fisher exponent, $\tau$, features similar properties to the one previously analyzed in 2D. One can conclude that the \new{set of exponents of a particular distribution} of points is free of \new{border} effects \new{(i.e., voids, a factor of particular importance in the quantification of the pair correlation function \cite{VillegasRS})}, and only depend on the underlying \new{intrinsic} properties of the point process.

\begin{figure}[hbtp]
\centering
\includegraphics[width=0.9\columnwidth]{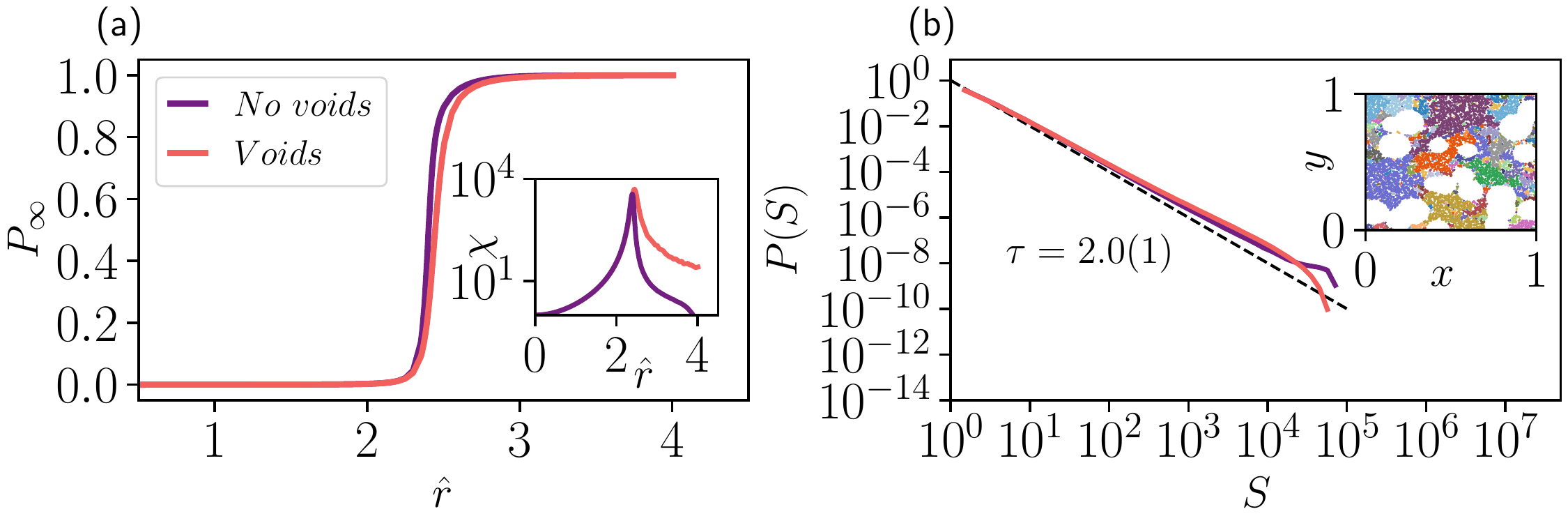}
\caption{\textbf{(A)} $P_{\infty}$ as a function of the distance parameter, $\hat r$, in terms
of the mean nearest-neighbor distance for $10^5$ random points with and without empty areas (see legend). There exists a phase transition at a critical value $\hat r_c\sim2.39(1)$ with identical qualitative features for both systems. Inset: Susceptibility for the two different cases. \textbf{(B)} Cluster size distribution, $P(S)$, versus size, $S$, with and without empty areas at the critical value, $\hat r_{c}$. Dashed line shows the \new{expected theoretical }exponent $\tau=2.05$. Inset: Cluster distribution of $N=10^4$ points in a system with empty areas at criticality.  Curves have been averaged over $10^{3}-10^{4}$
point patterns. \label{Voids}}

\end{figure}

\subsection{Inhomogeneous spatial point processes} \label{inhomogeneous1}

A heterogeneous Poisson process is defined by a set of non-interacting points, with intensity altered by external factors at different locations. Thus, the process is described through some predefined \emph{intensity function}, $\lambda(\mathbf{x})$, which depends on its spatial location $\mathbf{x}$. In the particular case of  tree species in tropical forests, inhomogeneous processes with density gradients are often influenced by topography or soil nutrients availability \cite{MoloneyPP}. We illustrate here two of the simplest non-homogeneous processes in 2D: (i) a Gaussian kernel around a ``central tree'', that is, $\lambda(r)={\displaystyle \frac{1}{\sqrt{2\pi\sigma^{2}}}e^{-\frac{r^{2}}{2\sigma^{2}}}}$ and, (ii) an exponential gradient along some predefined spatial direction, $\lambda(x)=le^{-lx}$.

\begin{figure}[hbtp]
\centering
\includegraphics[width=0.85\columnwidth]{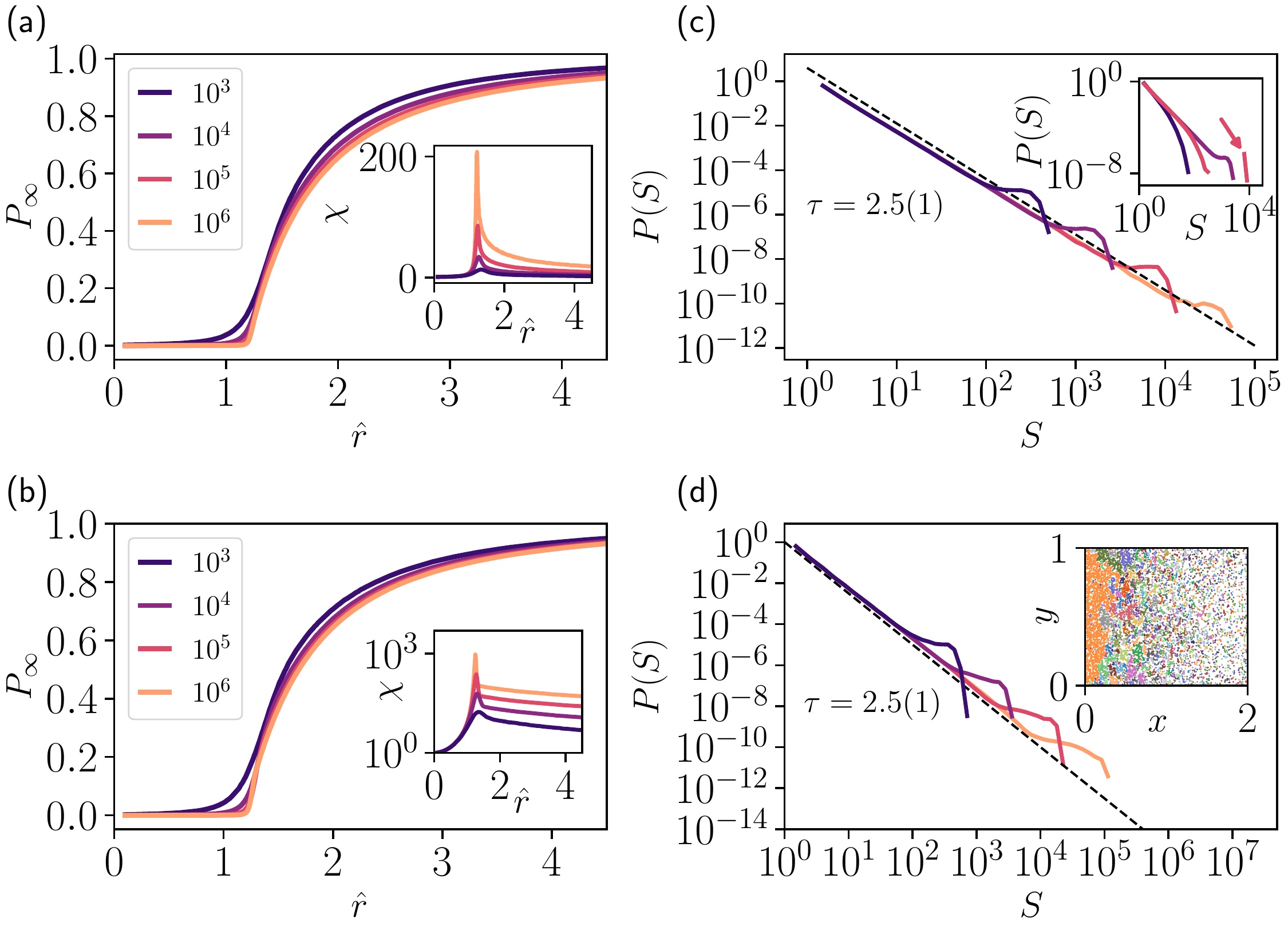}
\caption{$P_{\infty}$ as a function of the distance parameter, $\hat r$, in terms
of the mean nearest-neighbor distance for different number of points
(i.e. system size, see legend) in the case of: \textbf{(a)} a Gaussian intensity function with $\sigma=20$ and \textbf{(b)} an exponential density function with $l=1$. In both cases there exists a bonafide phase transition at a critical value $\hat r_c\sim1.2$ scaling with $N$. Inset: Susceptibility for different system sizes. \textbf{(c)} and \textbf{(d)} Cluster size distribution, $P(S)$, versus size, $S$, for different system sizes at the critical value, $\hat r_{c}$ for both cases, Gaussian and exponential, respectively. Dashed line shows the exponent $\tau=2.5(1)$. Insets: \textbf{(c)} $P(S)$ versus $S$ for the Gaussian case with $N=10^4$ points for a subcritical case ($\hat r = 0.8$), critical case ($\hat r_c = 1.26$) and supercritical case ($\hat r =2.0$). Observe that, for the supercritical case, the probability distribution shows a well-defined bump characteristic of a spanning cluster invading all the system.  \textbf{(d)} Cluster distribution of $N=10^4$ points in the exponenital case at criticality.  Curves have been averaged over $10^{3}-10^{4}$ point patterns. \label{GaussK}}

\end{figure}

Figures \ref{GaussK}(a) and (b) show the probability that a site belongs to the infinite cluster, $P_\infty$, versus the normalized distance $\hat r$ for both cases. Observe that in both situations, there exists a bonafide phase transition at $\hat {r}_c\sim1.2$. Figure \ref{GaussK}(c) and (d) show the power-law scaling, together with the characteristic exponent $\tau$. In such a case, the exponent presents an excellent convergence to the value $\tau=2.5(1)$. Interestingly, the present approach suggests an analogy between our percolation transition and a static percolation model known as gradient percolation (GP) \cite{Gab2001, Gab2000}. In particular, Figure \ref{GaussK} evidence the existence of an external frontier of the connected occupied cluster, which is often called the \emph{gradient percolation front}. Also, our results are in full agreement with recent simulations evidencing a size-distribution exponent for gradient percolation $\tau \sim 2.45$ \cite{Manna2022}, which might be indeed different from the mean-field expected one.

\subsection{Poisson cluster processes} \label{clustered}

Point processes with clustering are of utmost relevance for practical applications in ecology. They contrast areas of elevated density (i.e., clusters) with areas of low (or even vanishing) point density. For example, they have allowed to analyze the spatial distribution of the seedling process of the orchid $Orchis$ $purpurea$ \cite{jacquemyn2007} or reproduce empirical data for diverse species in Barro Colorado island \cite{Wiegand2009}.

We consider the simplest point process that generates clustered patterns with one critical scale of clustering: the Thomas process. It is defined by the following rules \cite{MoloneyPP}: (i) Consider $n_p$ 'parent' events following an homogeneous Poisson process. (ii) Each parent produces a fixed number of 'offspring,' $S$, thus being $N=n_pS$ the total system size. (iii) The offspring is seeded from the parent independently and identically distributed according to a radially symmetric normal distribution with variance $\sigma^2$.

We propose a brief heuristic argument for arguing the emergent phenomenology in this specific case. For the case of a small number of parents, we return to the case of the Gaussian intensity function presented above (thus expecting a single critical point scaling up to $S$ with exponent $\tau=2.5$). Otherwise, it will reduce to the usual two-dimensional case limit (i.e., $\tau=2.05$) when considering an infinite number of parents with almost any offspring. Nevertheless, there is, perhaps, an intermediate behavior. Let us consider that each center exhibits a characteristic dispersal radius, $2\sigma$, which will enclose 95$\%$ of the total number of children points. Thus, the percolation threshold of a 2D set of offspring disks of radius $\alpha$, for fully penetrable discs \cite{quintanilla2000}, should mark the beginning of fully covering the available space. Therefore, we can expect some anomalous emergent behavior around this critical number of parents,
\begin{equation}
n^c_p=\frac{\eta_c L_x L_y}{\pi(\alpha\sigma)^2},
\label{criticalTh}
\end{equation}
where $\eta_c\sim1.128$ represents the critical filling factor for fully penetrable discs \cite{quintanilla2000}, $L_x$ and $L_y$ are the dimensions of the filled area, $\sigma$ is the variance of the dispersal distance of the points and, as we have already said, we consider $\alpha\sim2$. Let us remark that it stresses the 'critical' condition for having a saturated environment as usually proposed: \textit{'large landscapes are essentially always biotically saturated with individuals'} \cite{hubbell2001}. 
\begin{figure}[hbtp]
\begin{centering}
\includegraphics[width=1.0\columnwidth]{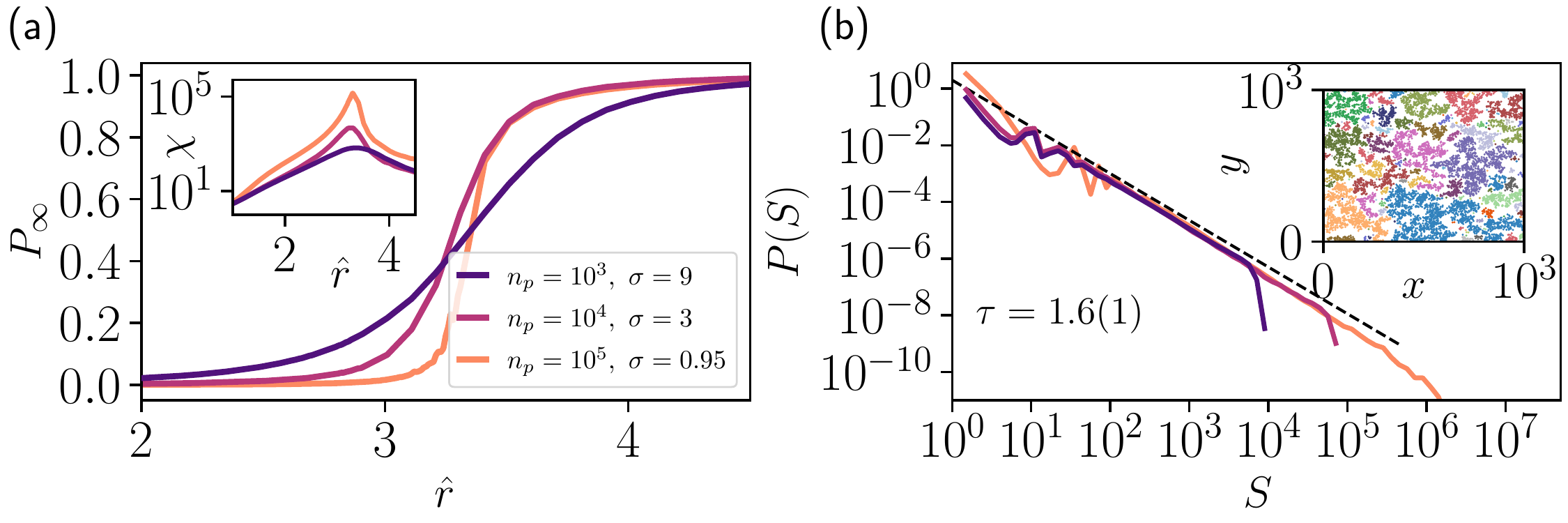}
\par\end{centering}
\caption{\textbf{(a)} $P_{\infty}$ as a function of the distance parameter $\hat r$ in terms
of the mean nearest-neighbor distance for different number of parents points and different dispersal distances (see legend). The offspring is fixed to $S=10$ in the two first cases and to $S=40$ in the last one. It is possible to observe a clear phase transition at
a critical value $\hat r_c\sim3.2$ scaling with $N$. Inset: Susceptibility
for the three different considered cases. \textbf{(b)} Cluster size distribution, $P(S)$ versus size, $S$, for the three selected cases at the critical value, $\hat r_{c}$. Dashed line shows the exponent $\tau=1.6(1)$. Inset: Cluster distribution of $n_p=10^3$ parent points and $S=10$ children points with dispersal radius $\sigma=9$.  Curves have been averaged over $10^{3}-10^{4}$
point patterns. \label{TProc}}
\end{figure}

Figure \ref{TProc}a shows the probability that a site belongs to the infinite cluster, $P_\infty$, versus the normalized distance $\hat r$ for different realizations of the Thomas process with a variable number of parents points in a squared area of size $L=10^3$, and scaling $\sigma$ and $n_p$ as suggested by Eq. \ref{criticalTh}. Figure \ref{TProc}(b) shows the cluster size distribution scaling, together with the characteristic exponent $\tau$. In this case, we observe a suitable convergence to the exponent $\tau=1.6(1)$, much smaller than expected for the isotropic percolation universality class. We highlight that it is compatible whether with mean-field directed percolation critical exponents \cite{LogPot}, invasion percolation ones \cite{Cafiero1996} or the emergence of holes in backbone percolation \cite{Hu2016}. To avoid confusion, let us underline that in this specific case, $\tau$ is smaller than the conventional lower bound value 2 and that power laws with exponents smaller than 2 can appear but do not have a well-defined averaged cluster size when integrated to arbitrarily large values of $S$. Therefore, the fits are just approximated ones and cannot possibly extend to arbitrarily large cluster sizes (we refer to \cite{Manna2022,Paula} for recent percolation numerical studies showing emergent exponents $\tau < 2$ and to \cite{Christensen2008} for an extended discussion on the issue). The phase transition reflects a characteristic system scale that somehow reflects the underlying clustering processes that give rise to the specific spatial landscape. 

\section{Conclusions}
In summary, we have confirmed that the aggregation of random point patterns in the continuum belongs to the isotropic percolation universality class \cite{gawlinski1981,vicsek1981,lee1990, Mertens2012} and can be extensively computed by carefully analyzing the scaling properties of the system as a function of the total number of points. Also, the particular consideration of density gradient automatically leads to the emergence of the gradient percolation universality class \cite{Gab2000,Gab2001}, reflecting the intrinsic heterogeneity of the point pattern structure. Our approach, \new{as expected for all local correlation functions that deal with the local random geometry of clusters}, allows us to confirm the ensemble equivalence in continuum percolation problems.

An interesting corollary is that some critical exponents (e.g., $\beta$) change --as a function of $\hat r$, in the spatial point aggregation problem-- if the spatial dimension is not explicitly included in the FSS analysis, and might lead to a misleading interpretation of the system's universality class. In particular, we have also shown that, in these specific circumstances (i.e., not considering the physical dimension of the system), analogous scaling relationship to those of complex networks apply \cite{Radicchi2010}. In particular, it is of potential interest to analyze further extensions in random geometric graphs \cite{Dall2002} and classify universality classes in random networks growth, where the spectral dimension can be easily controlled. In our opinion, our present work provides a new perspective to understand scaling dependences and universality classes on systems lacking a well-known spatial embedding by considering the spectral dimension of complex network structures, which can crucially constrain the emergent scaling properties of the system. 

Typically, point-process models are built to represent a hypothesized process that mimics natural spatial patterns, depending on pre-assigned statistical features based on the density field $\rho(\mathbf{x})$, the pair correlation function and its variations, or nearest-neighbors statistics \cite{MoloneyPP,andrea2004}. We demonstrate here that clustering of point-patterns reveals diverse percolative transitions that only depend on their intrinsic correlations and spatial properties, e.g., reflecting  \emph{homogeneity} or \emph{heterogenity}, together with density gradients and clustering properties. More importantly, here we show that this specific method is free of boundary and edge effects, which are essential for avoiding spurious behaviors in the analysis of density fields and pair correlation functions \cite{hanisch1984,VillegasRS,AShapesAC,MoloneyPP}. Consequently, this allows it to be used to extract fundamental information about real point patterns and their generative processes.  Additionally, we provide different values for the critical radius at which the percolation transition occurs --in complete agreement with the corresponding filling factors \cite{Torquato2012}-- that can be relevant, for instance, in ecological aggregation processes \cite{Scanlon, Kefi, Villegas2021}, wireless mobile {\em ad hoc} communication networks \cite{glauche2003}, or avalanche in bursty dynamics \cite{notarmuzi2021}.  Let us finally mention that detailed analyses of the particular aggregation of ecological landscapes are in progress and will be reported elsewhere. In our opinion, this contribution will help to clarify the study of continuum percolation processes allowing to extract information about empirical observations in natural systems such as tropical forests.

\ack
P.V. acknowledge financial support from the Spanish 'Ministerio de Ciencia e Innovaci\'on' and the 'Agencia Estatal de Investigaci\'on (AEI)' under Project Ref. PID2020-113681GB-I00. G.C. acknowledges the EU project 'HumanE-AI-Net', no. 952026. We also thank M.A.Mu\~noz, A.Maritan and S.Suweis for extremely valuable discussions and/or suggestions on earlier versions of the manuscript.

\appendix 

\section{Relation between $\eta_c$ and $\hat r_c$} \label{App0}

Let us consider the mean nearest neighbor distance of a random set of
$N$ points distributed in a $d-$dimensional space. In the hypothesis of a $d-$dimensional homogeneous Poisson point process \cite{GabrielliBook}, the PDF of the nearest-neighbor distances in a set of $N$ random points is given by the following expression:
\begin{center}
\begin{equation}
P_{d}\left(r\right)dr=\frac{2\pi^{\frac{d}{2}}}{\Gamma\left(\frac{d}{2}\right)}r^{d-1}dr\frac{N}{V}\left(1-\frac{\pi^{\frac{d}{2}}}{\Gamma\left(\frac{d}{2}+1\right)}r^{d}/V\right)^{N-1}
\end{equation}
\par\end{center}

Then, the mean nearest neighbor distance is
\begin{equation}
\left\langle r\right\rangle =\int_{0}^{\infty}rP_{d}\left(r\right)dr=\frac{2\pi^{\frac{d}{2}}}{\Gamma\left(\frac{d}{2}\right)}\rho\int_{0}^{\infty}r^{d-1}dr\left(1-\frac{\pi^{\frac{d}{2}}}{\Gamma\left(\frac{d}{2}+1\right)}r^{d}/V\right)^{N-1}
\end{equation}
which, under in the limit $N,V\to\infty$ with fixed $N/V=\rho$, becomes
\begin{equation}
\mbox{MNN}\equiv\left\langle r\right\rangle =\frac{2\pi^{\frac{d}{2}}}{\Gamma\left(\frac{d}{2}\right)}\rho\int_{0}^{\infty}r^{d}e^{-r^{d}\frac{\pi^{\frac{d}{2}}\rho}{\Gamma\left(\frac{d}{2}+1\right)}}dr=\frac{2\Gamma\left(\frac{1}{d}\right)\Gamma\left(\frac{d+2}{2}\right)}{\sqrt{\pi}\Gamma\left(\frac{d}{2}\right)d^{2}}\left[\frac{1}{\left[\frac{\rho}{\Gamma\left(\frac{d+2}{2}\right)}\right]^{\nicefrac{1}{d}}}\right]
\end{equation}
which can be inverted to give
\begin{equation}
\rho=\Gamma\left(\frac{d+2}{2}\right)\left[\frac{2\Gamma\left(\frac{1}{d}\right)\Gamma\left(\frac{d+2}{2}\right)}{\Gamma\left(\frac{d}{2}\right)d^{2}\mbox{MNN}\sqrt{\pi}}\right]^{d}\label{eq:MNN}
\end{equation}
The percolation threshold is characterized by the formation of a giant component which marks the phase transition. At this point the filling factor is defined by
\begin{equation}
\eta_{c}=\frac{\pi^{\frac{d}{2}}}{\Gamma\left(\frac{d}{2}+1\right)}r_{c}^{d}\rho\,,
\label{eq:Critical}
\end{equation}
where $r_c$ is the critical disk radius. 
Replacing Eq.(\ref{eq:MNN}) into Eq.(\ref{eq:Critical}), we can
get a direct relation between $\eta_{c}$ and the critical radius
--in terms of the $\mbox{MNN}$-- in the simple case of random points,
\begin{equation}
\hat{r}_{c}\equiv\frac{2r_{c}}{\mbox{MNN}}=d^{2}\frac{\Gamma\left(\frac{d}{2}\right)}{\Gamma\left(\frac{1}{d}\right)\Gamma\left(1+\frac{d}{2}\right)}\eta_{c}^{\nicefrac{1}{d}}\,,
\end{equation}
where \new{$d$ is the dimension of the system and} factor two is considered on the left-hand side to reflect
the fact that two disks will overlap if their center-to-center distance
is less than twice the radius of the disks. \new{Note that it reduces to $\hat r_c=4\sqrt{\nicefrac{\eta_c}{\pi}´}$ for $d=2$.}

\section{Periodic boundary conditions} \label{AppA}

We have extended the clustering algorithm proposed in the main text considering also periodic boundary conditions in two dimensions. Figure \ref{PBC} shows the comparison of the order parameter for both cases, with \new{(dashed line)} and without PBC \new{(solid line)}. Observe that there is no \new{substantial} difference between cases, \new{for large enough system sizes ($N>10^3$), thus confirming that we are in the limit of infinite lattice size where this effect becomes negligible}. We choose not to implement PBC for general simulations just for computational convenience.

\begin{center}
\begin{figure}[hbtp]
\begin{centering}
\includegraphics[width=0.75\columnwidth]{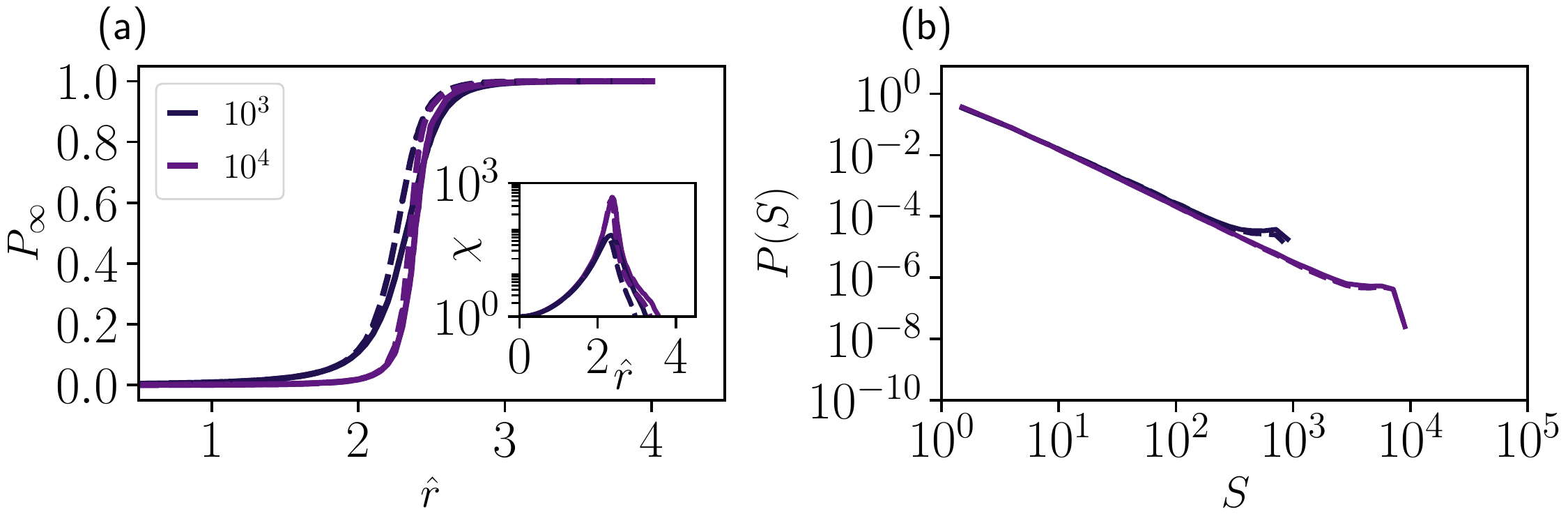}
\par\end{centering}
\caption{\textbf{(a)} $P_{\infty}$ as a function of the distance parameter $\hat r$ in terms
of the mean nearest-neighbor distance for different number of points
(i.e. system size, see legend) with periodic boundary conditions (dashed line) and without periodic boundary conditions (solid line). Inset: Susceptibility
for different system sizes in both situations. \textbf{(b)} Cluster size distribution, $P(S)$ versus size, $S$, for different system sizes at the critical value, $\hat r_{c}$. Dashed line shows the case considering periodic boundary conditions.  Curves have been averaged over $10^{3}-10^{4}$
point patterns. \label{PBC}}
 
\end{figure}
\par\end{center}

\section{Statistical ensembles} \label{AppB}

Our approach assumes two points belonging to the same cluster if their euclidean distance is less than or equal to $r$. In particular, we have considered a fixed number of points in space, i.e. system size $N$, which corresponds to a canonical statistical ensemble. There exists another different approach, which consists of considering an increasing number of discs, until some critical density $N/L^2$ is reached, generating a giant cluster in the system. Pay close attention to the fact that the last approach belongs to the grand-canonical ensemble.

\begin{center}
\begin{figure}[hbtp]
\begin{centering}
\includegraphics[width=0.5\columnwidth]{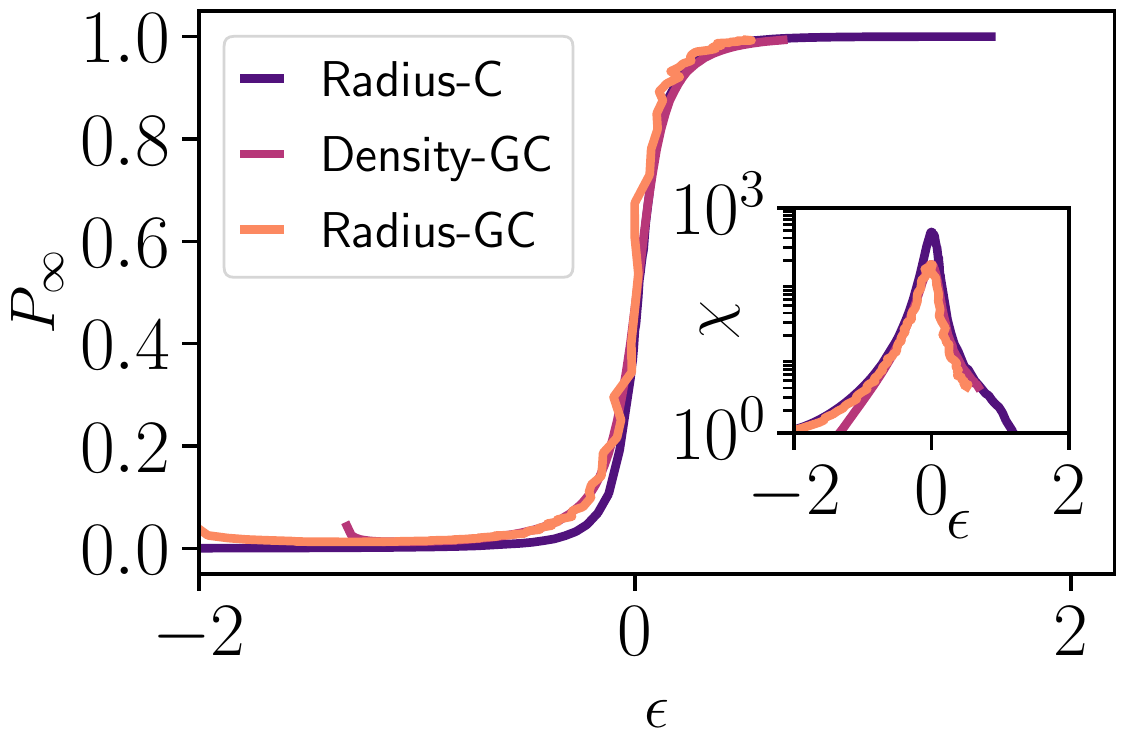}
\par\end{centering}
\caption{$P_{\infty}$ as a function of the distance to criticality $\epsilon$ for different ensembles (see legend). Inset: Susceptibility
for different statistical ensembles.  Curves have been averaged over $5\cdot10^{3}$ point patterns. \label{Ens}}

\end{figure}
\par\end{center}

Figure \ref{Ens} shows the results for the cluster strength, $P_\infty$, versus the distance to the critical point, $\epsilon\propto|\hat r-\hat r_c|$ in two different cases: (i) the canonical ensemble with $10^4$ points as done in the main text (RC, blue line in Fig.\ref{Ens}) and, (ii) the grand-canonical ensemble considering a square rectangle of size $L=50$ and adding an increasing number of discs with radius $r=1$. In this last case, it is possible to observe the phase transition using the density of discs as control parameter, $\rho=N/L^2$, where $\epsilon\propto|\rho-\rho_c|$ (D-GC, violet line in Fig.\ref{Ens}) or to calculate the mean nearest-neighbor distance, $\mbox{MNN}$ using the ratio between the radius of the discs and the mean nearest-neighbor distances as a control parameter, $\hat r\propto r/\mbox{MNN}$, where $\epsilon\propto|\hat r-\hat r_c|$ (R-GC, orange line in Fig.\ref{Ens}). Observe that, in all cases our results confirm the ensemble equivalence for large systems sizes, even if further analyses will be needed in the future to check, e.g., fluctuations for small systems \cite{Hu2012}.

\section{Beyond two dimensions} \label{AppC}

\subsection{Finite Size Scaling analysis}

We have computed the exponents $\nicefrac{\beta}{\nu d}$, $\nicefrac{\gamma}{\nu d}$, and $D$ for the different spatial dimensions. Figures \ref{Fit} and \ref{Fit1} show the different fits we have computed to obtain the critical exponents.

\begin{figure}[hbtp]
\centering
\includegraphics[width=0.75\columnwidth]{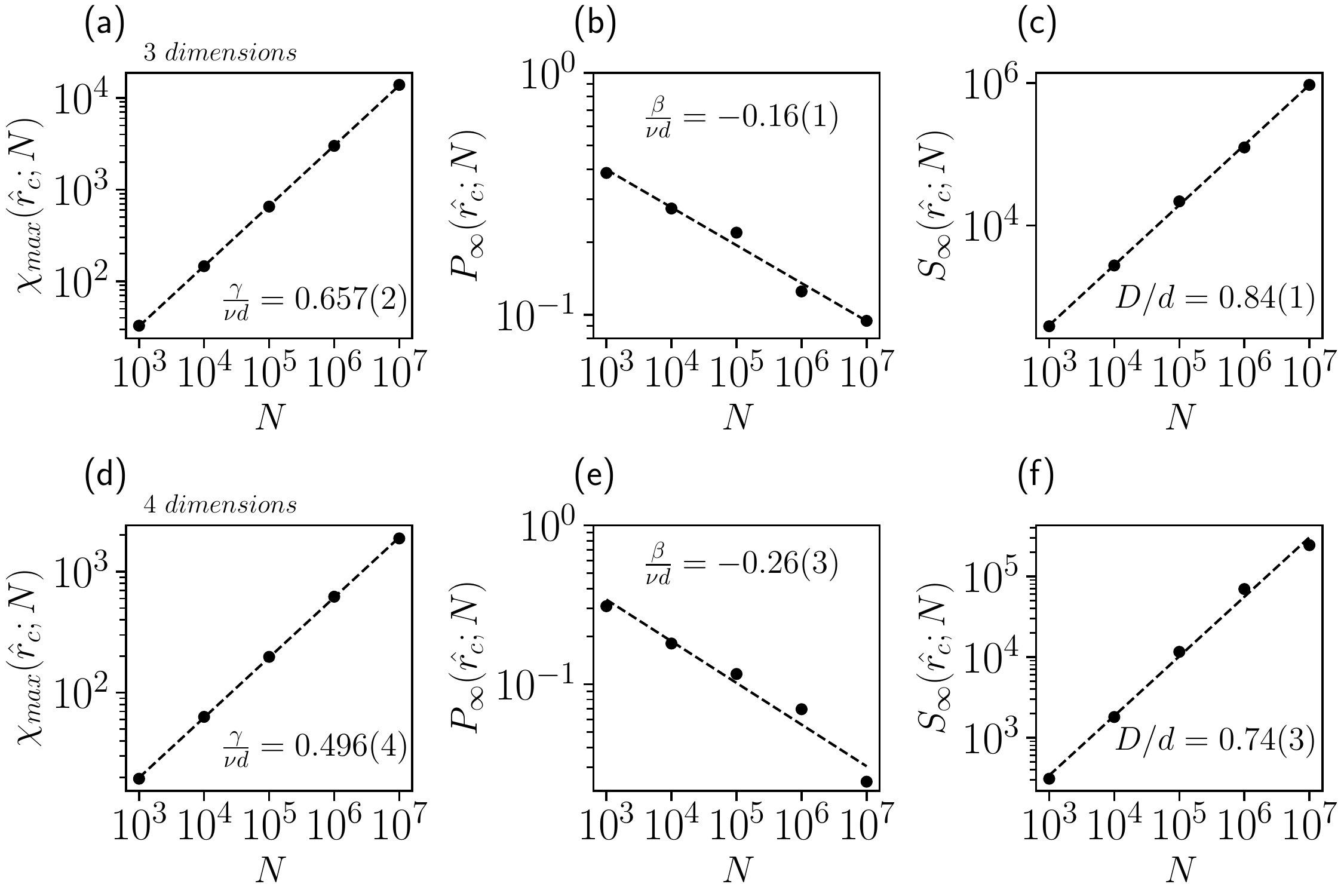}
\caption{\textbf{FSS analysis} for 3D and 4D. \textbf{(a,d)}  Scaling of the susceptibility at $\hat r=\hat r_c$, $\chi (\hat{r}_c;N) $, versus the system size, $N$. Black dashed line represent the best fit to our data, scaling as $\chi(\hat{r}_c;N)\propto N^{\gamma/\nu d}$. \textbf{(b,e)} Scaling of the order parameter at $\hat r=\hat r_c$, $P_\infty (\hat{r}_c;N) $, versus the system size, $N$. Black dashed line represent the best fit to our data, scaling as  $P_\infty(\hat{r}_c;N)\propto N^{-\beta/\nu d}$.  \textbf{(c,f)} Scaling of the maximum cluster at $\hat r=\hat r_c$, $S_\infty (\hat{r}_c;N) $, versus the system size, $N$. Black dashed line represent the best fit to our data, scaling as $S_\infty(\hat{r}_c;N)\propto N^{D/d}$. \label{Fit}}
\end{figure}

\begin{center}
\begin{figure}[hbtp]
\begin{centering}
\includegraphics[width=0.75\columnwidth]{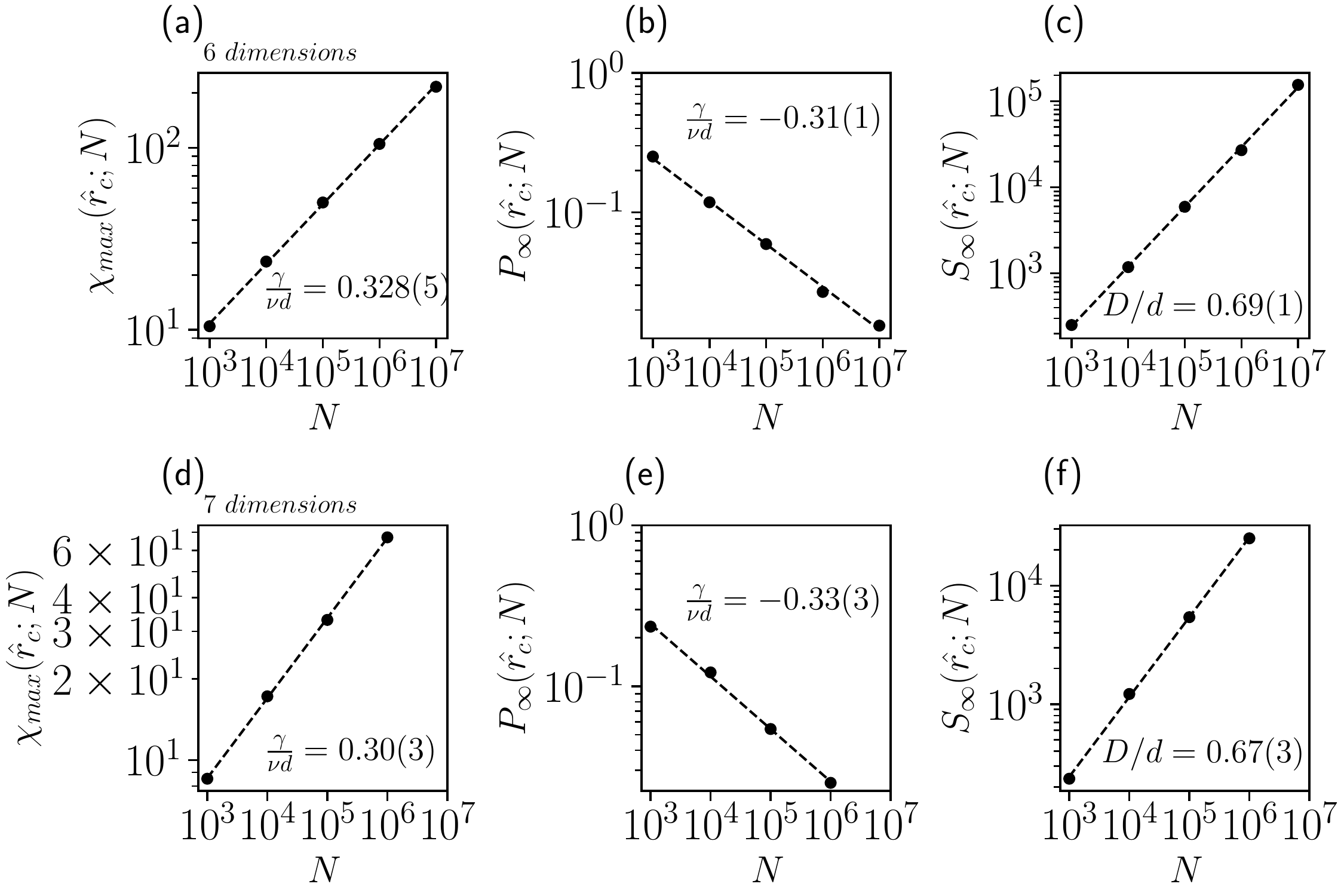}
\par\end{centering}
\caption{\textbf{FSS analysis} for 6D and 7D. \textbf{(a,d)}  Scaling of the susceptibility at $\hat r=\hat r_c$, $\chi (\hat{r}_c;N) $, versus the system size, $N$. Black dashed line represent the best fit to our data, scaling as $\chi(\hat{r}_c;N)\propto N^{\gamma/\nu d}$. \textbf{(b,e)} Scaling of the order parameter at $\hat r=\hat r_c$, $P_\infty (\hat{r}_c;N) $, versus the system size, $N$. Black dashed line represent the best fit to our data, scaling as  $P_\infty(\hat{r}_c;N)\propto N^{-\beta/\nu d}$.  \textbf{(c,f)} Scaling of the maximum cluster at $\hat r=\hat r_c$, $S_\infty (\hat{r}_c;N) $, versus the system size, $N$. Black dashed line represent the best fit to our data, scaling as $S_\infty(\hat{r}_c;N)\propto N^{D/d}$. \label{Fit1}}
\end{figure}
\par\end{center}

\newpage
Figure \ref{FSS0} and \ref{FSS1} show the FSS collapse employing the obtained set of critical exponents.

\begin{center}
\begin{figure}[H]
\begin{centering}
\includegraphics[width=1.0\columnwidth]{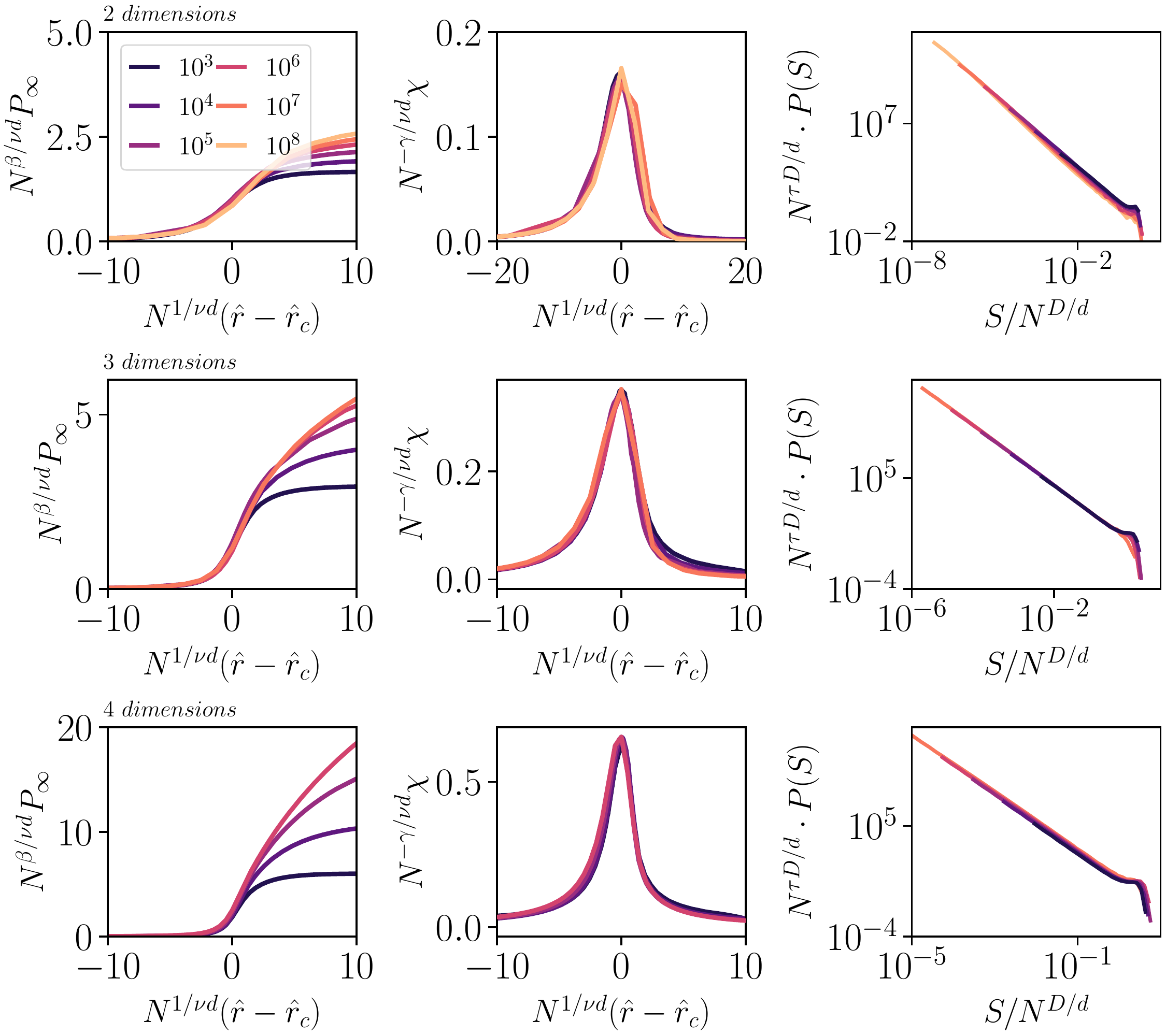}
\par\end{centering}
\caption{\textbf{(Left)} FSS collapse for the order parameter. \textbf{(Center)}  FSS collapse for the susceptibility. \textbf{(Right)} Data collapse analysis of the avalanche size distribution for different system sizes. \label{FSS0}}
 
\end{figure}
\par\end{center}
\newpage
\begin{center}
\begin{figure}[H]
\begin{centering}
\includegraphics[width=0.85\columnwidth]{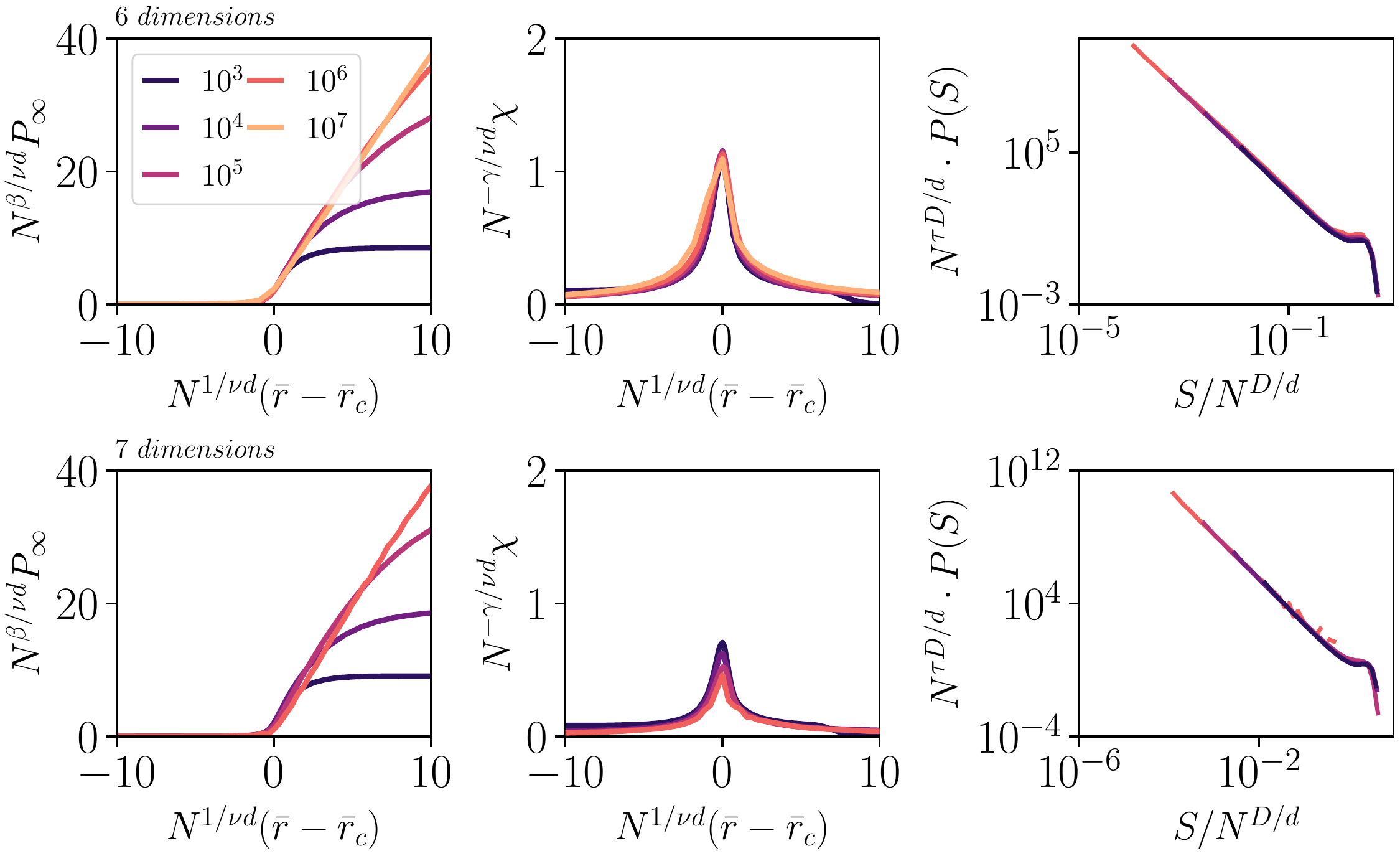}
\par\end{centering}
\caption{\textbf{(Left)} FSS collapse for the order parameter. \textbf{(Center)}  FSS collapse for the susceptibility. \textbf{(Right)} Data collapse analysis of the avalanche size distribution for different system sizes. \label{FSS1}}
 
\end{figure}
\par\end{center}

\section*{References}

\providecommand{\newblock}{}
\providecommand{\url}[1]{{\tt #1}}
\providecommand{\urlprefix}{}
\providecommand{\href}[2]{#2}
\def\url#1{}

\end{document}